\documentclass[12pt]{article}
\usepackage{amssymb,amsmath,amsthm}

\theoremstyle{plain}
\newtheorem{thm}{THEOREM}
\newtheorem{lem}{LEMMA}
\theoremstyle{definition}
\newtheorem{rem}{Remark}

\newcommand{\A}{{\mathsf A}}
\newcommand{\B}{{\mathsf B}}
\newcommand{\sA}{{\mathsf a}}
\newcommand{\sB}{{\mathsf b}}
\newcommand{\C}{{\mathbb C}}
\newcommand{\M}{{\mathsf M}}
\newcommand{\N}{{\mathbb N}}
\newcommand{\R}{{\mathbb R}}
\newcommand{\s}{{\mathcal S}}
\newcommand{\Tr}{{\rm Tr\, }}
\newcommand{\id}{{\mathbb I}}
\newcommand{\la}{{\lambda}}

\begin{document}
%\markboth{\scriptsize{August 5, 2002}}{\scriptsize{August 5,
%2002}}

\title{Equivalent forms of the Bessis-Moussa-Villani conjecture}
\author{\vspace{5pt} Elliott H.~Lieb$^{*}$ and Robert Seiringer$^{
\dagger}$\\
\vspace{-4pt}\small{ Departments of Mathematics and Physics,
Jadwin Hall,} \\
\small{Princeton University, P.~O.~Box 708, Princeton, New Jersey
  08544}}
\date{\small January 17, 2003}
\maketitle

\renewcommand{\thefootnote}{$*$}
\footnotetext{Work partially
supported by U.S. National Science Foundation
grant PHY 01 39984.}
\renewcommand{\thefootnote}{$\dagger$}
\footnotetext{Erwin Schr\"odinger Fellow,
supported by the Austrian Science Fund.\\
\copyright\, 2003 by the authors. This paper may be reproduced, in its
entirety, for non-commercial purposes.}

{\it Dedicated with best wishes to Giovanni Jona-Lasinio on his
70$^{th}$ birthday }

\begin{abstract}
The BMV conjecture for traces, which states that $\Tr \exp(\A -\la \B)$ is the
Laplace transform of a positive measure, is shown to be equivalent to
two other statements: (i) The polynomial
$\lambda\mapsto\Tr(\A+\lambda\B)^p$ has only non-negative coefficients for
all $\A, \B \geq 0$, $p\in \N$ and (ii) $\lambda\mapsto \Tr (\A+\lambda
\B)^{-p}$ is the Laplace transform of a positive measure for $\A, \B
\geq 0$, $p>0$.
\end{abstract}

%\section{Introduction}
An intriguing conjecture \cite{bessis}, which is more than a quarter
century old, was formulated by Bessis, Moussa and Villani (BMV) in an
attempt to simplify the calculation of partition functions of
quantum mechanical systems.  It refers to a positivity property of
traces of matrices which, if true, would permit the calculation of
explicit error bounds in a sequence of approximations known as the
Pad\'e approximants.

The BMV conjecture is easy to state. Let $\A$ and $\B$ be any two Hermitian
$n\times n$ matrices with $\B \geq 0$, and let $\la \in \R$. Then the 
function of $\la$ given by
\begin{equation}
\la \mapsto \Tr \exp(\A -\la \B)
\end{equation}
(with $\Tr =$ trace) is the Laplace transform of a positive measure on
$[0,\infty)$.

Oddly, the positivity can be easily seen to be true for quantum
mechanics defined by the Schr\"odinger equation for bosons without
magnetic fields because the partition function in that case can be
represented by Wiener integrals. This representation fails for
fermions, and hence the importance of the conjecture for condensed
matter physics.

The BMV conjecture appears naturally in other areas of matrix analysis
as well and there is a significant literature devoted to it. For
$2\times 2$ matrices there is an easy proof of its correctness, but
for $3\times 3$ matrices neither a proof nor a counterexample is known
to this day.  Some recent work that shows the validity of the
conjecture in some `average' or `typical' sense is in
\cite{petz2,petz1}. Ref. \cite{moussa} contains a review. 

The purpose of the present paper is to show that the BMV conjectured
positivity is equivalent to two other positivity statements that
appear to be somewhat different. One of them is, in fact, easier than
the original conjecture because it involves only a statement about
certain polynomials.  Namely for every pair of positive matrices $\A$ and $\B$ and every positive integer $p$, the polynomial
$\lambda \mapsto \Tr (\A+\lambda \B)^p$ has all its coefficients
positive.  It is relatively easy to see that this statement implies
the BMV conjecture but the striking fact, which is the main point of
our paper, is that the BMV conjecture implies the polynomial
statement.  The current consensus is that the BMV conjecture is true,
in which case our equivalence leads to a non-obvious and interesting
theorem about traces.  It also presents an alternative route to
proving or disproving the BMV conjecture. We are indebted to D. Petz
for urging us to write this note about the equivalences.

\begin{thm}\label{T1}
For fixed $n$ 
let $\A$ and $\B$ denote arbitrary hermitian $n\times n$ matrices over $\C$,
and let $\lambda\in\R$. 
The following statements are equivalent:
\begin{itemize}
\item[(i)] For all $\A$ and $\B$ positive, and all $p\in\N$, the polynomial
$\lambda\mapsto\Tr(\A+\lambda\B)^p$ has only non-negative coefficients.
\item[(ii)] For all $\A$ hermitian and $\B$ positive, $\lambda\mapsto\Tr
\exp{(\A-\lambda\B)}$ is the Laplace transform of a positive
measure supported in $[0,\infty)$.
\item[(iii)] For all $\A$ positive definite and $\B$ positive,
and all $p \geq 0$,  $\lambda\mapsto \Tr (\A+\lambda \B)^{-p}$ is the
Laplace transform of a positive measure supported in $[0,\infty)$.
\end{itemize}
\end{thm}

\begin{rem} 
$\A$ positive means that $\A$ is Hermitian and $(v,\A v) \geq 0$ for
all vectors $v$. Positive definite means positive and invertible.
\end{rem}

\begin{rem} By Bernstein's theorem \cite{don}, a function of $\la$ defined for
$\la \geq 0$ is the Laplace transform of a positive measure supported
in $[0,\infty)$ if and only if it is completely monotone, i.e., its
$r$-th derivative has the sign $(-1)^r$ for all values of $\la\geq 0$.
\end{rem}

\begin{rem} In the case of $2\times 2$ matrices, it is easy to see
that statement (i) of Theorem~\ref{T1} (and therefore all three
statements) are true: there exists a basis in which both $\A$ and
$\B$ have only non-negative entries, and therefore all coefficients of
$\Tr(\A+\lambda \B)^p$, which is independent of the choice of the
basis, are necessarily non-negative. Some other special cases are 
treated in \cite{kumar}.
\end{rem}

\begin{rem} Note the logical structure in the proof that 
statement (iii) for $p \in \N$ implies item (i), which then
implies (iii) for all $p>0$.
\end{rem}

\begin{rem} The proof will show (see (\ref{ltran})) 
that item (iii) is really a very simple corollary of item
(ii). Indeed, the proof extends to the statement that $\la \mapsto \Tr
f(\A+\la \B)$ is the Laplace transform of a positive measure supported
in $[0,\infty)$ for $\A\geq 0, \B\geq 0$ whenever $f$ is the Laplace transform of a positive
measure supported in $[0,\infty)$.
\end{rem}

\begin{rem} 
An important remark about item (i) is addressed in
\cite{johnson}. When $\Tr(\A+\lambda\B)^p$ for $p\in \N$ is
multiplied out, the coefficient of $\la^r$ is a sum of
terms. \cite{johnson} shows that some of these terms can be
negative. The first interesting case occurs for $p=6$ and $r=3$. The
term $\Tr \A^2 \B^2 \A \B$ can be negative even though $\A $ and 
$\B$ are positive.  Nevertheless, the BMV
conjecture implies that the sum of all terms of order $\la^3$ is
non-negative. We cannot prove this, even for $n=3, r=3, p=6$.
\end{rem}

The proofs of (i)$\Rightarrow$(ii) and (ii)$\Rightarrow$(iii) will
turn out to be very easy. The difficult direction is (iii)$\Rightarrow$(i)
and the following lemma is needed for this case.

\begin{lem}\label{L1}
Let $\sA$ and $\sB$ be hermitian $n\times n$ matrices over $\C$,
with $\sA$ positive definite. Define $\A=\sA^{-1}$
and $\B=\sA^{-1/2}\sB \sA^{-1/2}$, and let $\lambda\in\R$. For all
$p,r\in \N$
\begin{equation}\label{equiv}
\left.\frac {d^r}{d\lambda^r} \Tr \frac 1{\left(\sA+\lambda
\sB\right)^{p}}\right|_{\lambda=0}=\frac{p}{p+r} (-1)^r \left.
\frac {d^r}{d\lambda^r} \Tr \left(\A+\lambda
\B\right)^{p+r}\right|_{\lambda=0} \ .
\end{equation}
\end{lem}

\begin{proof}
By induction it is easy to show that
\begin{equation}
\frac {d^r}{d\lambda^r} \left(\A+\lambda \B\right)^{p+r}=r!
\sum_{\substack{0\leq i_1,\dots,i_{r+1}\leq p \\ \sum_j i_j=p }}
(\A+\lambda \B)^{i_1} \B \cdots \B (\A+\lambda\B)^{i_{r+1}} \ .
\end{equation}
By taking the trace at $\lambda=0$ we obtain
\begin{equation}
I_1\equiv \left. \frac {d^r}{d\lambda^r} \Tr \left(\A+\lambda
\B\right)^{p+r}\right|_{\lambda=0}=r! \sum_{\substack{0\leq
i_1,\dots,i_{r+1}\leq p
\\ \sum_j i_j=p }} \Tr \A^{i_1} \B \cdots \B
\A^{i_{r+1}}   \ .
\end{equation}
Moreover, by similar arguments,
\begin{equation}
\frac {d^r}{d\lambda^r} \frac 1{\left(\sA+\lambda
\sB\right)^{p}}=(-1)^r r! \sum_{\substack{1\leq
i_1,\dots,i_{r+1}\leq p
\\ \sum_j i_j=p+r }} \frac 1{(\sA+\lambda \sB)^{i_1}} \sB \cdots \sB
\frac 1{(\sA+\lambda\sB)^{i_{r+1}}} \ .
\end{equation}
By taking the trace at $\lambda=0$ and using cyclicity, we get
\begin{equation}
I_2\equiv\left.\frac {d^r}{d\lambda^r} \Tr \frac
1{\left(\sA+\lambda \sB\right)^{p}}\right|_{\lambda=0}=(-1)^r r!
\sum_{\substack{0\leq i_1,\dots,i_{r+1}\leq p-1
\\ \sum_j i_j=p-1 }} \Tr \A\, \A^{i_1} \B \cdots \B
\A^{i_{r+1}}\, \ .
\end{equation}
We have to show that
\begin{equation}
I_2=\frac{p}{p+r} (-1)^r I_1 \ .
\end{equation}
To see this we rewrite $I_1$ in the following way. Define $p+r$
matrices $\M_j$ by
\begin{equation}
\M_j=\left\{\begin{array}{ll}\B & {\rm for\ }1\leq j\leq r \\ \A &
{\rm for\ } r+1\leq j\leq r+p\ .\end{array}\right.
\end{equation}
Let $\s_n$ denote the permutation group. Then
\begin{equation}
I_1=\frac 1{p!} \sum_{\pi\in\s_{p+r}} \Tr \prod_{j=1}^{p+r}
\M_{\pi(j)} \ .
\end{equation}
Because of the cyclicity of the trace we can always arrange the
product such that $\M_{p+r}$ has the first position in the trace.
Since there are $p+r$ possible locations for $\M_{p+r}$ to appear in the
product above, and all products are equally weighted,  we get
\begin{equation}
I_1=\frac {p+r}{p!} \sum_{\pi\in\s_{p+r-1}} \Tr \A
\prod_{j=1}^{p+r-1} \M_{\pi(j)} \ .
\end{equation}
On the other hand,
\begin{equation}
I_2=(-1)^r \frac 1{(p-1)!} \sum_{\pi\in\s_{p+r-1}} \Tr \A
\prod_{j=1}^{p+r-1} \M_{\pi(j)}\ ,
\end{equation}
so we arrive at the desired equality.
\end{proof}

\begin{proof}[Proof of Theorem \ref{T1}]
(i)$\Rightarrow$(ii): By continuity of the trace,
\begin{equation}
\Tr \exp{(\A-\lambda \B)} = e^{-\|\A\|}\sum_{k=0}^\infty \frac
1{k!} \Tr(\A+\|\A\|\, \id -\lambda \B)^k \ ,
\end{equation}
where $\id$ denotes the identity matrix. 
It follows from Bernstein's theorem and (i) that the right side is the
Laplace transform of a positive measure supported in $[0,\infty)$.

(ii)$\Rightarrow$(iii): This follows from taking the trace of the
matrix equation
\begin{equation}\label{ltran}
\frac1{(\A+\lambda\B)^p}=\frac 1{\Gamma(p)} \int_0^\infty
\exp \left[-t(\A+\lambda \B)\right] t^{p-1} dt \ .
\end{equation}

(iii)$\Rightarrow$(i): It suffices to assume (iii) only for $p\in
\N$. For invertible $\A$ we observe that the $r$-th derivative of $\Tr
(\sA+\lambda \sB)^{-p}$ at $\la=0$ is related to the coefficient of
$\la^r$ in $\Tr(\A+\lambda\B)^p$ as given by (\ref{equiv}) with $\A,
\sA, \B, \sB$ related as in Lemma \ref{L1}.  The left side of
(\ref{equiv}) has the sign $(-1)^r$ because it is the derivative of a
function that is the Laplace transform of a positive measure supported
in $[0,\infty)$.  Thus the right side has the correct sign as stated
in item (i). The case of non-invertible $\A$ follows from continuity.
\end{proof}

\end{document}